\documentclass[runningheads,a4paper]{llncs}
\usepackage{cite}
\usepackage{amssymb} 
\usepackage{graphicx}
\usepackage{url}
\usepackage{float}
\usepackage{soul,color}  
\usepackage{amsmath}
\usepackage{todonotes}
\usepackage{dirtytalk}
\usepackage{adjustbox}
\usepackage[inline]{trackchanges}

\begin{document}

\newcommand{\luiz}[1]{\textcolor{red}{#1}}

\title{From DevOps to DevDataOps: Data Management in DevOps processes}

\author{Antonio Capizzi \inst{1}, Salvatore Distefano \inst{1},Manuel Mazzara \inst{2}\\ 
\institute{University of Messina, Italy
\and Innopolis University, Innopolis, Respublika Tatarstan, Russian Federation.
}}

\toctitle{Lecture Notes in Computer Science}
\tocauthor{Authors' Instructions}

\authorrunning{Capizzi, Distefano, Mazzara}

\maketitle

\abstract{
DevOps is a quite effective approach for managing software development and operation, as confirmed by plenty of success stories in real applications and case studies. DevOps is now becoming the main-stream solution adopted by the software industry in development, able to  reduce the time to  market  and costs while improving  quality  and ensuring  evolvability and adaptability of the resulting software architecture. Among the aspects to take into account in a DevOps process, data is assuming strategic  importance,  since  it allows to gain  insights from the operation directly into the development,  the main objective of a DevOps approach. 
Data can be therefore considered as the fuel of the DevOps process, requiring proper solutions for its management. 
Based on  the  amount  of data generated,  its  variety,  velocity, variability, value and other relevant  features,  DevOps  data  management can be mainly framed into the BigData category. 
This allows exploiting BigData solutions for the management of DevOps data generated throughout the process, including artefacts, code, documentation, logs and so on. 
This paper aims at investigating data management in DevOps processes, identifying related issues, challenges and potential solutions taken from the BigData world as well as from new trends adopting and adapting DevOps approaches in data management, i.e. DataOps.
}   


\section{Introduction}

DevOps \cite{Bass,8314153} is an approach for software development and (IT) system operation combining best practices from both such domains to improve the overall quality of the software-system while reducing costs and shortening time-to-market.
Its effectiveness is demonstrated by the quite widely adoption of DevOps approaches in business contexts, where there is a big demand of specific professionals such as DevOps engineers as well as data scientists, just partially, minimally covered by current offer.   

The DevOps philosophy can be generalized as a way, a good practice for improving a generic product or service development and operation, by connecting these through a feedback from operation to development.
An important feature of DevOps is the automation of such a process:  \textit{continuous delivery} (CD) enables organizations to deliver new features quickly and incrementally by implementing 
a  flow of changes into the production via an automated 
``assembly line" - the \textit{continuous delivery pipeline}. 
This is coupled with \textit{continuous integration} (CI) that aims at automating the software/product integration process of codes, modules and parts, thus identifying a CI/CD pipeline.

The tools adopted to implement this high degree of automation in the DevOps process identifies a \textit{toolchain}. 
DevOps toolchain tools are usually encapsulated into different, independent containers deployed into physical or virtual servers (typically on Cloud), and then managed by specific scripts and/or tools (e.g. Jenkins), able to orchestrate and coordinate them automatically.

Such DevOps principles have been therefore either specialized to some specific software/application domains (security - SecOps, SecDevOps, DevSecOps \cite{devsecops}, system administration - SysOps  \cite{sysops}, Web - WebOps or WebDevOps \cite{webops}) or even adopted, rethought and adapted in other contexts such as artificial intelligence (AIOps \cite{aiopsdevops}) and machine learning (MLOps, DeepOps \cite{mlops}), and data management (DataOps \cite{dataops}).  
The latter, DataOps, aims at mainly organizing data management according to DevOps principles and best practices. 
To this end, DataOps  introduces the concept of \textit{dataflow pipeline} and toolchain, to be deployed in containerized (Cloud) environment providing feedback on performance and QoS of the overall data management process, used to real-time tune  the pipeline to actual operational needs and requirements.

As discussed above, the DevOps pipeline automation 
involves in the toolchain different tools, 
each continuosly generating messages, logs  and data including artifacts.
To achieve DevOps aims and goals, such data has to be properly managed, collected, processed and stored to provide insights from operations to the development stages. 
DevOps data management could therefore be quite challenging, due to the large amount of data to be considered as well as its variety, variability and similar metrics usually identified as V properties in the BigData community, to which we have to refer to.
BigData approaches, indeed, could be a good solution to consider in the management of a DevOps process and toolchain.


In light of these considerations, in this paper we focus on DevOps data management, proposing to adopt BigData approaches and solutions.
More specifically, the main goal of this paper is to explore the convergence between DevOps and DataOps approaches, defining a possible (big)dataflow pipeline for DevOps processes and toolchains and organizing it following a DataOps process, towards DevDataOps.
This way, we investigate on the adoption of DataOps, mainly implementing a BigData pipeline and toolchain, in DevOps contexts, i.e. for improving the development and operation of a software architecture.

To this extent, Section \ref{sec:devdataops} describes the DevOps and DataOps processes and toolchains.
Section \ref{sec:data} discusses about DevOps artifacts and data in the BigData context.
Then, Section \ref{sec:ddopl} proposes a DevOps (big) dataflow pipeline and related implementation in the DataOps philosophy.
Section \ref{sec:conclusions} summarises the key aspects of the proposed approach and future work.



\section{DevOps and DataOps}
\label{sec:devdataops}
\subsection{The DevOps process and toolchain}
\label{sec:devops}
DevOps \cite{Bass} consists of a set of practices to promote collaboration between the developers, IT professionals (in particular sysadmin, i.e. who works on IT operations) and quality assurance personnel. 
DevOps is implemented via a set of software tools \cite{8314153} that enable the management of an environment in which software can be built, tested and released quickly, frequently, and a more reliable manner. 
In addition to \textit{continuous delivery} (CD), which aims at developing ``small'' software releases in reasonably short cycles, \textit{continuous integration} (CI) stands as a key concept in DevOps approaches.
A typical example of CI consists of continuously integrating changes made by developers into a repository, then a project build is automatically executed, and if the build the modifications are integrated into the code through CD and published in the production environment.

A DevOps process is usually composed of different  stages and phases, which can be periodically reiterated for proper development and operation of the software architecture, in an evolutionary fashion able to caught new requirements, features and behaviors arising from operation. 
This way, a DevOps process belogs to the category of agile process, not plan driver, where the number of development cycle is unknown a-priori. 
Consequently, the amount of data generated by a DevOps process is usually unpredictable and could be really high.
There is no standard definition of a DevOps process, but several different versions and implementations have been provided by the related community. 
Among them, main DevOps stages can be summarized below.

\begin{itemize}
    \item \textit{Plan}: activity planning and task scheduling for the current release. This step is usually dealt with by project managers in collaboration with the  team  and exploiting project management tools such as Trello, Asana, Clarizen, Jira, Azure DevOps, to name a few.

\item \textit{Code}: code development and code review. Developers are the most closely involved in this activity using IDE and tools for source code management such as GitHub, Artfactory, CodeClimate, etc. 

\item \textit{Build}: is when source code is converted into a stand-alone form that can be run on a computer system. In this activity are involved various professional figures, mainly developers and sysadmins. The tools used in this phase are: CI tools (for example Jenkins, TravisCI), build tools (for example Maven) etc.

\item \textit{Test}: in this phase the software is tested by the quality assurance staff using tools for (automatic) testing and similar.  
Examples of such kind of tools are JUnit, Jmeter, Selenium, etc.

\item \textit{Release}: triggered when a new version of software is ready to be released to end users. In this activity, various professionals of the development team are involved, primarily developers and sysadmins. 
Release management tools (such as Spinnaker) or similar support such an activity.

\item \textit{Deploy}: it deals with the installation and execution of the new software release in the production environment and infrastructure. 
At this stage, the collaboration between developers and syadmins is mandatory. 
The tools used for deployment depend on the  target infrastructure (physical or virtual nodes, Cloud, etc.) as well as on the  adopted system software (OS, virtualization, containerization, middleware, compilers, libraries), thus identifying a wide set of possible options for deployment  management (VmWare, Virtualbox, Docker, LXD, Kubertenes, AWS CodeDeploy, ElasticBox etc.).

\item \textit{Operate}: is the activity that maintains and adapts the infrastructure in which the software is running. 
This activity is mainly driven by  sysadmins, supported by configuration management tools (such as Ansible, Puppet, Chef), security tools (such as Sonarqube, Fortify SCA, Veracode), database management tools (such as Flyway, MongoDB), recovery tools (PowerShell, Ravello), etc.

\item \textit{Monitor}: in this activity the software in production is monitored by mainly sysadmins, operators and others managing the project. 
The tools used are: tools that monitor the performance of the service, tools that analyze the logs (for example Logstash, Nagios, Zabbix), tools that analyze the end user experience (Zenoss).

\end{itemize}

One of the main objectives of DevOps is to mitigate issues in production, which is done by reducing the gap among development and testing environments to the production one. 
To this purpose,  
several tools as the one mentioned above are usually used and combined into
a set identified as the ``DevOps Toolchain", which can be considered as a scaffold built around the development project.
To compose a Toolchain, in general, there are no fixed rules, it is necessary to follow the DevOps principles and best practices to choose the tools according to the project characteristics. 
For a small project 3 or 4 tools might be enough, while in a larger project 10 or more tools might be necessary.
A minimal (CI/CD) DevOps toolchain might include, at least, some version control tool (e.g. Git), automation tools (e.g. Jenkins), package managers (e.g. NPM) and test tools (e.g. JUnit). 
DevOps infrastructures are typically 
fully implemented on Cloud platforms. 
It is a good practice in DevOps to build the entire infrastructure using containers to minimize portability issues. 
For that, containerization technologies  such as Docker, LXD or similar are adopted, sometimes coupled by tools for containers orchestration (e.g. Kubernetes or Swarm).

\subsection{DataOps}
\label{sec:dataops}
DataOps is a new approach that aims to improve quality and responsiveness of data analytics life-cycle \cite{dataops1,dataops2, dataops}.
This approach is based on DevOps rules, in particular DataOps aims to bring DevOps benefits to data analytics, adopting Agile rules and Lean concepts.
When the volume of data is larger and larger, the purpose of DataOps is to improve the life cycle of analytics by taking advantage of DevOps  principles such as communication between teams (data scientists, ETL, analysts, etc.), cooperation, automation, integration, etc. 
To achieve this it is necessary to apply a set of human practices and dedicated tools.
With DataOps a new professional figure called ``DataOps Engineer" was born, to deal with the automation and orchestration of the process.
A large DataOps community issued a Manifesto\footnote{https://www.dataopsmanifesto.org/} that contains 18 rules, the mission and best practices to apply DataOps.
However, this is the most concrete activity behind DataOps, that is still mainly a set of rules, concepts and ideas to be applied to data management.
There is, indeed, a lack of examples, dataflow pipelines, toolchians and standard process for DataOps.

However, despite the DataOps approach is still mainly abstract, it can be mostly summarized with the DataOps principles detailed in the Manifesto, some implementations of the DataOps idea start to be defined and fixed.
For example, a DataOps process can be broadly organized into three steps \footnote{https://dzone.com/articles/dataops-applying-devops-to-data-continuous-dataflo}:
\begin{itemize}
    \item \textit{Build} - In this step, the data is taken from a source point (e.g. a database, a log file, etc.), transformed by applying one or more actions, and then written to a destination point. The flow executed by these actions is called ``dataflow or DataOps pipeline". 
    In the Build phase you can also have multiple pipelines connected to each other.
\item \textit{Execute} - at this stage the build pipelines are put into production in a running environment e.g. clusters, datacenters, Clouds. 
It is important, for a company adopting DataOps, to be able to use the existing infrastructure to run the pipeline, in order to avoid incurring in additional costs.
\item \textit{Operate} - When this step is reached, the system is running on an environment, it is necessary to monitor it and react to any change (for example when larger volumes of data arrive and it is required to scale the infrastructure to cope with burst). One approach used in DataOps (borrowed from Agile) is to start with small instances and increase their resources when the demand grows.
\end{itemize}


\section{DevOps data}
\label{sec:data}
An outcome from the complex pipeline involved in a DevOps project is the generation of a large amount of data, considering the process artefacts and the log files generated in each stage. 
Examples of activities that generate considerable data on the project cycle include changes made by developers; the application building and its corresponding entries on the compilation and dependencies of the project; the execution of automatic tests;  software usage by end-users after release into the production.
Furthermore, the number of cycle of a DevOps process is usually unknown, and typically lasting years, so reaching hundreds, or even thousands of releases each generating a considerable amount of data that should be adequately preserved and managed for gaining insights and the process. 

Artifacts and data produced in a DevOps process are quite large and widely different. 
These can include software artifacts (code, documentation, test, executable, prototypes) and other information generated by the DevOps toolchain (logs, configuration files, traces, ...).
More specifically, based on the above DevOps reference process, the data associated to each stage is reported below.

\begin{itemize}
    \item \textit{Plan}: planning artifacts and data are software design blueprint, requirement documentations (UML or similar, if any), project environment information including user stories, tasks, activities, backlog, and statistics.

\item \textit{Code}: development artifacts include codes, versions, prototypes and related info such as lines of code, version differences and relevant parameters.

\item \textit{Build}: mainly executable files, packages, logs and metrics that contain information about builds and may indicate compilation errors, warnings, successes, failures etc.

\item \textit{Test}: code for automatic tests, logs from automatic test tools that indicates unit tests failed or passed, system tests results, or documentation written by Quality Assurance staff about verification on software in development.

\item \textit{Release}: documentation about releases (for example new features introduced), new final version of executable files or packages, logs and metrics from release orchestration tools.

\item \textit{Deploy}: configuration files, scripts and logs originating from Deploy tools that may contain errors or warnings.

\item \textit{Operate}: data generated by the software, logs from the tools involved in this stage and system logs from (physical or virtual) servers.

\item \textit{Monitor}: logs, metrics and other information from monitoring tools, the data retrieved in this phase is important to obtain a feedback from users.
\end{itemize}

It could be worth to invest in a data management system for
a DevOps process, where the high volume of generated data is not only properly collected and stored but also managed, filtered, aggregated and possibly made available for further processing,
to gain insights on the overall process to achieve essential DevOps aims and goals.
This calls for proper data management techniques, providing mechanisms for collection, aggregation, storage 
filtering, aggregation, fusion, archival, mining and feature extraction, local and global analytics preferably in an automated manner \cite{Protasov2018},
to improve the DevOps pipeline \cite{Kontogiannis:2018}. 
For example, historical data can be analyzed to estimate a probabilistic measure of the success of a new release, or for identifying potential source of bugs (root-cause analysis) or even to prevent them.

From a data/information-oriented perspective therefore, a DevOps process can be considered as a data-intensive process, in the sense that it could generate large amount of data.
To this purpose, DevOps data management issues and challenges can be framed into the BigData context.
Considering the reference DevOps process and toolchain described in Section \ref{sec:devops}, it could be interesting to characterize such a process in BigData terms.
To this purpose, we refer to well-known and widely used Bigdata metrics: the ``Vs".
BigData V properties are usually used in the community to categorize an application, and range in number from original 3 (Volume, Velocity, Variety) to 10 or even more.
Shortly, \textit{volume} is probably the best known characteristic of big data, quantifying the amount of data generated;
\textit{velocity} refers to the speed at which data is being generated, produced, created, or refreshed;
\textit{variety} is related to the ``structuredness" of data: we don't only have to handle structured data (logs, traces, DB) but also semistructured and mostly unstructured data (images, multimedia files, social media updates) as well; 
\textit{variability} refers to 
inconsistencies in the data,
to the multitude of data dimensions resulting from multiple disparate data types and sources and  
to the inconsistent speed at which big data is loaded into DB;
\textit{veracity} is the confidence or trust in the data,
mainly referring  to the provenance or reliability of the data source, its context, and how meaningful it is to the analysis based on it;
\textit{validity}
refers to how accurate and correct the data is for its intended use;
\textit{vulnerability} is concerned with big data security, privacy and confidentiality;
\textit{volatility} refers to data ``lifetime", i.e. the amount of time needed for data  to be considered irrelevant, historic, or not useful any longer;
\textit{visualization}
faces technical challenges due to limitations of in-memory technology and poor scalability, functionality, and response time to 
represent big data (billion data points) such as data clustering or using tree maps, sunbursts, parallel coordinates, circular network diagrams, or cone trees;
\textit{value} is the property to be 
derived 
from the data through processing and analytics. In the DevOps context, the process data value is exploited 
to support decision making in development.
This way,  Table \ref{tab:devopsbd} reports the characterization of a DevOps pipeline from a BigData perspective, expressed in terms of V metric values ranges for a ``mid-size" DevOps reference process.

\begin{table}[ht!]
\centering
\begin{adjustbox}{width=\textwidth}
\begin{tabular}{|c|c|c|c|c|c|c|c|c|c|c|}
\hline
\textbf{\textit{Stage/Vs}} &\textit{Volume} & \textit{Velocity} & \textit{Variety} & \textit{Variability} & \textit{Veracity} & \textit{Validity} & \textit{Vulnerability}& \textit{Volatility} & \textit{Visualization}& \textit{Value} \\ \hline \hline 
\textbf{Plan} & 10KB-1GB& Week& UnStr. & Medium/High & High & Low &   Low &Week/Days &Poor&High\\ \hline 
\textbf{Code} & 1-100MB & Hours & SemiStr. & High & High & High & Medium & Hours &Poor&High\\ \hline 
\textbf{Build} & 1-10GB & Hours & SemiStr & Medium & Low & Low & Medium & Hours &High &High \\ \hline 
\textbf{Test}& 10KB-1GB & Minutes & Str & Medium & High & High & Medium & Days & High &Medium\\ \hline 
\textbf{Release}& 1-10GB & Week & UnStr & High & Medium & Medium & Medium & Week/Month & Medium & High\\ \hline 
\textbf{Deploy}& 1-100MB & Week & UnStr & High & Medium & Medium & Medium & Week/Month & Medium & High\\ \hline 
\textbf{Operate} & 10KB-1GB & Hours & SemiStr & High & High & High & Medium & Hours & Medium & High\\ \hline 
\textbf{Monitor}& 10KB-1GB & Seconds/Minutes & SemiStr & High & High & High & High & Hours & High & High\\ \hline 
\end{tabular}
\end{adjustbox}
\caption{DevOps Project Bigdata Vs.}
\label{tab:devopsbd}
\end{table}



\section{DevDataOps}
\label{sec:ddopl}

\begin{figure}[ht!]
  \centering
  \includegraphics[scale=0.30]{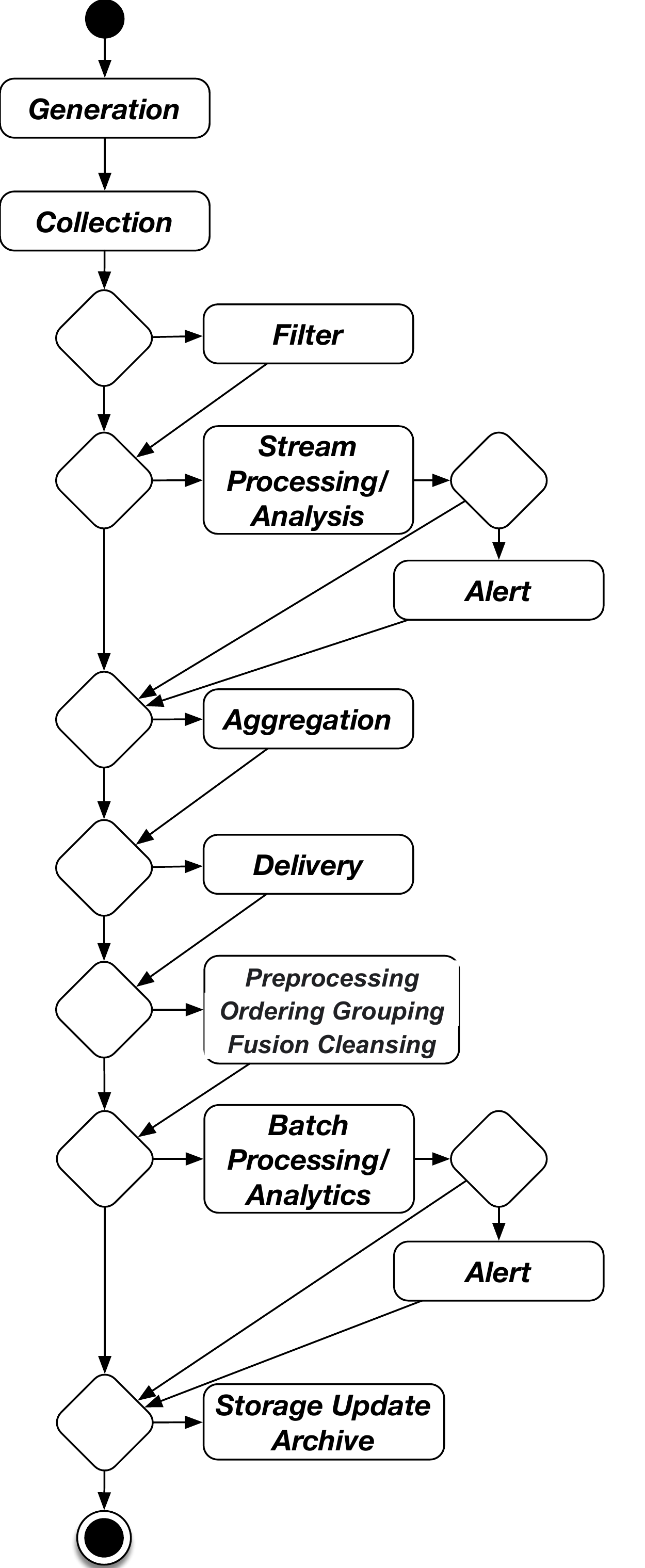}
  \caption{DevOps dataflow pipeline} 
  \label{fig:logs-mgmt-wflow}
\end{figure}

\subsection{DevOps Dataflow Pipeline}
\label{sec:ddoppl}
The data generated by a DevOps pipeline is therefore quite complex and heterogeneous, and consequently quite hard to manage and  maintain \cite{Chen19}. 
The DevOps data life-cycle and workflow can be decomposed into different stages and steps identifying the dataflow pipeline 
shown in Fig. \ref{fig:logs-mgmt-wflow}.
This could be considered as a quite generic DataOps pipeline that can be generally applied to any DevOps process, after adaptation, represented in Fig. \ref{fig:logs-mgmt-wflow} by the conditional diamonds modeling the presence or absence of a specific step in the DevOps dataflow-DataOps pipeline. 


\begin{itemize}
    \item \textbf{Generation}: Each module of the DevOps pipeline generates a log stream reporting its operation through a specific monitoring process.
    
    \item \textbf{Collection}: The generated logs are collected, reordered according to their timestamps, and grouped altogether to provide a snapshot of the whole DevOps process for each time interval.
    
    \item\textbf{Filtering}: Logs are then filtered to remove outliers, replicas, or observations that may contain errors or are undesirable for analysis. Logs filters are usually based on temporal statistics, i.e. based on previously logs average or similar statistical moments.
    

    \item \textbf{Stream processing, analysis and alerting}: The processing of single logs streams is locally performed in nearly real-time to identify potential flaws, defects or errors in a particular DevOps pipeline stage. In such cases, warning, error messages or activities are triggered by the alerting tool.
    
    \item \textbf{Aggregation}: The logs are aggregated and expressed in a summary form for statistical analysis. The main goal of the aggregation is to compress the volume of data.
    
    \item \textbf{Delivery}: The logs are made available to end-users and applications. This data is typically transmitted through networks using related protocols to physical and virtual servers.
    

    \item \textbf{Preprocessing, ordering, grouping, cleansing, and fusion}: The logs are reordered according to their timestamp and grouped to provide a snapshot of the DevOps process at each time interval. Next, the logs have to be cleaned, that is, having redundancies removed and being integrated with different sources into a unified schema before storage. The schema integration has to provide an abstract definition of a consistent way to access the logs without having to customize access for each log source format. Still during the preprocessing stage, logs undergo through a fusion process aiming to integrate multiple sources to produce a more consistent and accurate information.


    \item  \textbf{Storage, update and archiving}: This phase aims the efficient storage and organization as well as a continuous update of logs as they become available. Archiving refers to the long-term offline storage of logs. The core of the centralized storage is the deployment of structures that adapt to the various data types and the frequency of the capture (e.g. relational database management systems).
    
    \item \textbf{Processing and analytics}: Ongoing retrieval and analysis operations on stored and archived logs, mainly offline for root cause analysis, to identify any correlation among stages, predict behaviors and support decisions. Analytics is the discovery, interpretation, and communication of meaningful patterns in data in the logs and can be performed at different levels with different objectives: \textit{descriptive} (what happened), \textit{diagnostic} (why something happened), \textit{predictive} (what is likely to happen) and \textit{prescriptive} (what action to take).

\end{itemize}

\subsection{DataOps implementation}
\label{sec:ddoppl}


The DevOps dataflow pipeline should be then implemented according to a DataOps approach, as reported below
\paragraph{Build - }
To implement the DevOps dataflow pipeline of Fig. \ref{fig:logs-mgmt-wflow} in a DataOps fashion, following the process described in Section \ref{sec:dataops}, we have to start with building the toolchain, thus identifying the tools associated with each of the pipeline step. 
In this case, the real benefit of adopting BigData solutions in DataOps is clearly manifested: this way, the DevOps dataflow pipeline of Fig. \ref{fig:logs-mgmt-wflow} can be just considered as a BigData workflow to be deployed exploiting a tool among the plethora available   to manage BigData workflow (Hadoop, Spark, Storm, Flink, Samza, NiFi, Kafka, NodRed, Crosser.io, to name but a few).
Mainly belonging to the Apache (big) family, they allows to define and manage BigData workflows composed of different tasks or processes, highly customizable and configurable, then linked through specific mechanisms and tools able to enforce the workflow topology, or even to further parallelize and optimize it.
\paragraph{Execute}
Once built as a BigData workflow, the DataOps toolchain needs to be deployed and executed.
At this stage, therefore, automation tools such as Jenkins and deployment tools such as Docker or Jupiter could be used to support and further automate the BigData one, by for example containerising tasks or connecting them with tools  (monitoring) external to the BigData workflow.
Usually, the target deployment  infrastructure for a DataOps toolchain is the Cloud, public as Microsoft Azure and Amazon EC2, or private such as those managed locally by OpenStack or similar middleware.
\paragraph{Operate}
The operation stage is mainly tasked at providing a feedback on the DataOps toolchain to the DataOps engineers that have to tune it based on this feedback.
Both the process and the underlying infrastructure running the toolchains have to be monitored.
Metric of interest to be benchmarked in this case could be system parameters (CPU, memory, storage utilization), process non functional properties (response time, reliability, availability, energy consumption), or even specific V properties (volume, velocity, variety, etc, see Section \ref{sec:data}) 
In this step, tools for monitoring such as Prometheus or Nagios, and for managing the infrastructure such as Chef, Puppet etc. can be exploited.

\section{Conclusions}
\label{sec:conclusions}

DevOps is a modern approach to software development aiming at accelerating the build lifecycle via automation. Google, Amazon and many other companies can now release software multiple times per day making the old concept of ``release" obsolete. Data is at the centre of all the DevOps process and requires BigData solutions to be managed. Despite of the growing importance of DevOps practices in software development, management of the data generated by the toolchain is still undervalued, if not entirely neglected. 

In this paper we investigated this cutting-edge aspect of software development identifying related issues, challenges and potential solutions. Solutions do not need to be entirely new since large literature has been already published in the field of BigData. This emerging field of research is often referred to as \textit{DataOps}. While DevOps was created to serve the needs of software developers, DataOps users are often data scientists or analysts.

The current work has just started scratching the surface of such a complex subject, and in the limited space could not explore all the detailed aspects of analytics. One of the project on which our team is working at the moment is the analysis of data generated by the DevOps toolchain in order to identify anomalies in the incoming releases \cite{Capizzi19}. The same idea can be applied to Microservices monitoring, and this is exactly our next goal towards NoOps.

\bibliographystyle{ieeetr} 
\bibliography{biblio}

\end{document}